%% file: MPQPN.tex
\documentclass[aps,pra,showpacs,amssymb,nofootinbib,superscriptaddress,twocolumn,longbibliography]{revtex4-1}
\usepackage[T1]{fontenc}
\usepackage[english]{babel}
\usepackage[utf8]{inputenc}
\usepackage[colorlinks,breaklinks]{hyperref}
\usepackage{float}
\usepackage{tabularx}
\usepackage{graphicx}
\usepackage{amsthm}
\usepackage{bbm}
\usepackage{tikz}
\usepackage[toc,page]{appendix}
\usepackage{titlesec}
\usepackage{xcolor}
\makeatletter
\newcommand\org@hypertarget{}
\let\org@hypertarget\hypertarget
\renewcommand\hypertarget[2]{%
  \Hy@raisedlink{\org@hypertarget{#1}{}}#2%
  }
\makeatother
\hypersetup{
	bookmarksnumbered,
	pdfstartview={FitH},
	citecolor={darkgreen},
	linkcolor={darkred},
	urlcolor={darkblue},
	pdfpagemode={UseOutlines}}
\definecolor{darkgreen}{RGB}{50,190,50}
\definecolor{darkblue}{RGB}{0,0,190}
\definecolor{darkred}{RGB}{238,0,0}

\usepackage[framemethod=tikz]{mdframed}
\definecolor{mycolor}{RGB}{12, 104, 180}
\definecolor{mycolor2}{RGB}{168, 188, 204}
\newmdenv[innerlinewidth=0.5pt, roundcorner=4pt,linecolor=mycolor,innerleftmargin=6pt,
innerrightmargin=6pt,innertopmargin=6pt,innerbottommargin=6pt]{mybox}

\usepackage{paracol}
\usepackage{tcolorbox}
\newcommand\tabletitlefontsize{\fontsize{9pt}{11pt}\selectfont}

\tcbset{colback=mycolor2!5 , colframe=mycolor, fonttitle=\bfseries\tabletitlefontsize,float=htb}
\newtcolorbox[blend into=figures]{boxfigure}[3][]
{ float*=ht,width=\textwidth,lower separated=false, center upper,
title={#2},label= fig:#3,#1}
\newtcolorbox[blend into=figures]{smallboxfigure}[3][]
{float=ht,lower separated=false, blend before title=colon hang,
title={#2}, label= fig:#3 ,#1}
\newtcolorbox{smallbox}[3][]
{float=ht,lower separated=false, blend before title=colon hang,
title={#2}, label= fig:#3 ,#1}
\newtcolorbox[blend into=tables]{smallboxtable}[3][]
{float=tb,lower separated=false, blend before title=colon,left=7pt,
title={#2}, label=table:#3 ,#1}
\newtcolorbox[blend into=tables]{bigboxtable}[3][]
{float*=t,lower separated=false, blend before title=colon hang, width = 2\linewidth,
title={#2}, label= table:#3 ,#1}

\newtcolorbox[blend into=tables]{sideboxtable}[3][]
{float*=hbt,sidebyside, sidebyside align = top, righthand width=.38\textwidth,  lower separated=false, blend before title=colon hang, width = 2\linewidth,
title={#2}, label= table:#3 ,#1}

\newcolumntype{Z}{|>{\centering\arraybackslash}X}

\usepackage{MnSymbol}
\usepackage{enumerate}%allows different styles of enumerate environment
\hypersetup{
	bookmarksnumbered,
	pdfstartview={FitH},
	citecolor={darkgreen},
	linkcolor={darkred},
	urlcolor={darkblue},
	pdfpagemode={UseOutlines}}
\definecolor{darkgreen}{RGB}{50,190,50}
\definecolor{darkblue}{RGB}{0,0,190}
\definecolor{darkred}{RGB}{238,0,0}
\usepackage{soul}
\usepackage{mathtools}

\newcommand{\be}{\begin{equation}}
\newcommand{\ee}{\end{equation}}
\newcommand{\ben}{\begin{equation*}}
\newcommand{\een}{\end{equation*}}
\newcommand{\bea}{\begin{eqnarray}}
\newcommand{\eea}{\end{eqnarray}}

\newcommand{\tr}{\textnormal{Tr}}

\newcommand{\ket}[1]{\ensuremath{\left|\right.\!{#1}\!\left.\right\rangle}}

\newcommand{\ketbra}[2]{\ensuremath{|{#1}\rangle\langle{#2}|}}

\newcommand{\djj}{d\kern-0.4em\char"16\kern-0.1em}

\definecolor{magenta}{rgb}{1.0, 0.0, 0.56}

\usepackage{multirow}

\begin{document}

\title{A Large-Scale Reconfigurable Multiplexed Quantum Photonic Network}

\author{Natalia Herrera Valencia}
    \email[Email address: ]{n.herrera_valencia@hw.ac.uk}
    \affiliation{Institute of Photonics and Quantum Sciences, Heriot-Watt University, Edinburgh, UK}

\author{Annameng Ma}

\author{Suraj Goel}

\author{Saroch Leedumrongwatthanakun}
    \altaffiliation[Current address: ]{Division of Physical Science, Faculty of Science, Prince of Songkla University, Songkhla, Thailand}
    
\author{Francesco Graffitti}
    \altaffiliation[Current address: ]{Cyberhawk, Edinburgh, UK}

\author{Alessandro Fedrizzi}
    
\author{Will McCutcheon}
    
\author{Mehul Malik}
    \email[Email address: ]{m.malik@hw.ac.uk}
    \homepage[Website: ]{http://bbqlab.org}
    \affiliation{Institute of Photonics and Quantum Sciences, Heriot-Watt University, Edinburgh, UK}

%\date{\today}

\begin{abstract}

Entanglement distribution in quantum networks will enable next-generation technologies for quantum-secured communications, distributed quantum computing and sensing. Future quantum networks will require dense connectivity, allowing multiple users to share entanglement in a reconfigurable and multiplexed manner, while long-distance connections are established through the teleportation of entanglement, or entanglement swapping. While several recent works have demonstrated fully connected, local multi-user networks based on multiplexing, extending this to a global network architecture of interconnected local networks remains an outstanding challenge. Here we demonstrate the next stage in the evolution of multiplexed quantum networks: a prototype global reconfigurable network where entanglement is routed and teleported in a flexible and multiplexed manner between two local multi-user networks composed of four users each. At the heart of our network is a programmable $8\times 8$-dimensional multi-port circuit that harnesses the natural mode-mixing process inside a multi-mode fibre to implement on-demand high-dimensional operations on two independent photons carrying eight transverse-spatial modes. Our circuit design allows us to break away from the limited planar geometry and bypass the control and fabrication challenges of conventional integrated photonic platforms. Our demonstration showcases the potential of this architecture for enabling large-scale, global quantum networks that offer versatile connectivity while being fully compatible with an existing communications infrastructure.

\end{abstract}
\maketitle

%%%%%%%%%%%%%%%% INTRO %%%%%%%%%%%%%%%%%%%%%%%

%Entanglement-based quantum networks
Quantum networks enable the distribution and processing of quantum information between distant interconnected quantum nodes, with applications ranging from quantum communication to distributed quantum computing~\cite{Kimble:2008uv}. The distribution of entanglement over multi-user quantum network architectures in an efficient and scalable manner is of paramount importance. In addition to enabling next-generation quantum communication technologies such as device-independent quantum key distribution~\cite{Vazirani:2014hu}, entanglement-based networks allow for a distributed configuration that can bolster the performance of quantum computation~\cite{Cirac:1999jr,Eisert2000} and quantum metrology~\cite{komar2014,Guo2019}.
 
\begin{figure*}[t!]
    \centering
    \includegraphics[width=\linewidth]{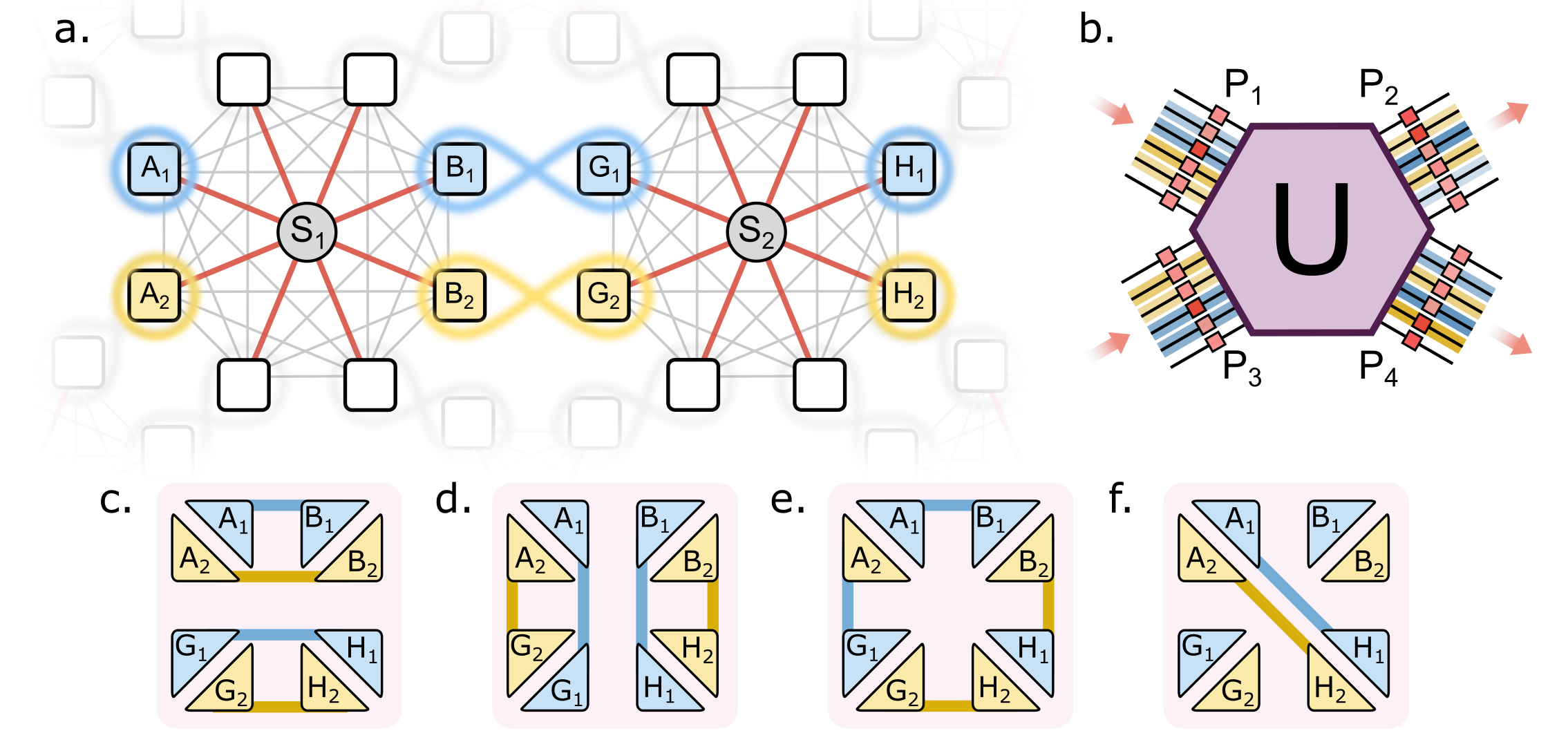}
    \caption{\textbf{a.~A global, multiplexed quantum network architecture.} Two local, multiplexed quantum networks consisting of entanglement sources (S$_1$ and S$_2$) and sets of four users (\{A$_1$, A$_2$, B$_1$, B$_2$\} and \{G$_1$, G$_2$, H$_1$, H$_2$\}) are connected to form a global, eight-user multiplexed quantum network that can route and swap entanglement in a reconfigurable manner. \textbf{b.~Programmable multi-port design.} A complex medium (U) is placed between four programmable phase planes (P$_1$-P$_4$) to implement an $8\times 8$-dimensional multi-mode circuit that operates on two independent photons to implement entanglement routing and swapping operations. \textbf{c-f.~Entanglement structures.} The global network can be reconfigured on-demand to realise different network structures connecting pairs of users with qubit-entanglement in several configurations, including multiplexed entanglement swapping between pairs of distant users A$_1$H$_1$ and A$_2$H$_2$. 
    }
    \label{fig:NetStruct}
\end{figure*}

%State of the art 
Realising a network where entanglement can be shared between multiple users requires flexible and high-capacity platforms for generating and distributing entanglement. State-of-the-art network implementations have demonstrated the flexible routing of bipartite entanglement between multiple users via hybrid multiplexing in bulk and integrated platforms \cite{Erhard:2017gl,Wengerowsky2018,Joshi2020,Appas2021,Alshowkan2022,Zheng2023}. These demonstrations harness an auxiliary photonic degree-of-freedom (for e.g.~wavelength) to multiplex several bipartite entanglement channels to realise fully connected, \textit{local} quantum networks of up to eight users. The natural next step in the evolution of such networks is to interconnect multiple, local, multi-user networks to realise a \textit{global} multiplexed network where distant users located at the edges of their respective local networks can share entanglement with each other (see Fig.~\ref{fig:NetStruct}a). A key obstacle holding back the realisation of such a global network is the lack of reconfigurable network devices capable of simultaneously routing and teleporting/swapping entanglement over multiplexed channels. As the number of photons and modes processed by such a device increase, maintaining device performance while minimising coupling losses becomes a significant challenge.

Conventionally, such network devices have been realised using the ``bottom-up'' approach, which relies on a planar mesh arrangement of a large number of carefully controlled 2-mode interferometers~\cite{Reck_1994,Clements2016}. While impressive dimensionalities of up to 20 modes in an integrated photonic circuit have been demonstrated recently \cite{Taballione2023}, increasing circuit complexity further will require overcoming many challenges in circuit design and fabrication. Thermal management of the large number of components limits current performance and hinders further scalability~\cite{Kyatam2019}. An additional hurdle is presented by the difficulty of efficiently coupling single photons from free-space or fibre links into an integrated circuit.

%Relevance for the multiplexed platform: complex media circuits
In this context, complex scattering media such as multi-mode fibres (MMFs) have recently emerged as promising candidates for both the transport and manipulation of quantum states of light~\cite{Defienne:2016dk, Leedumrongwatthanakun:2019wt}. Light scattering inside an MMF can be described as a unitary and random linear optical process in a very high-dimensional space. While this effect can be detrimental to the correlations of an entangled state, it can be reversed via knowledge of the fibre transmission matrix \cite{Popoff:2010cj}, enabling the transport of high-dimensional spatial entanglement through a complex channel~\cite{Valencia2020}. 
Using inverse-design techniques, the light scattering process inside a complex medium can also be harnessed to realise high-dimensional programmable optical circuits by placing the medium (U) between controllable phase planes (P) (see Fig.~\ref{fig:NetStruct}b). This alternative ``top-down'' approach simplifies the circuit architecture by separating the control layer from the mixing layer by employing a large number of auxiliary modes, achieving fully programmable high-dimensional quantum gates \cite{goel2022inverse}. Furthermore, it allows one to break out of the restrictive planar geometry of photonic integrated circuits and access the full transverse-spatial photonic degree-of-freedom distributed volumetrically. Finally, using waveguides for both transporting and manipulating quantum states of light bypasses the problem of coupling losses between optical links and photonic chips. In recent work, we demonstrated how an MMF-based circuit can be used to manipulate a single photon in its high-dimensional spatial structure, and function as a generalized, programmable multi-outcome measurement device for manipulating and certifying high-dimensional entanglement~\cite{goel2022inverse}.

\begin{figure*}[ht!]
    \centering
    \includegraphics[width=\linewidth]{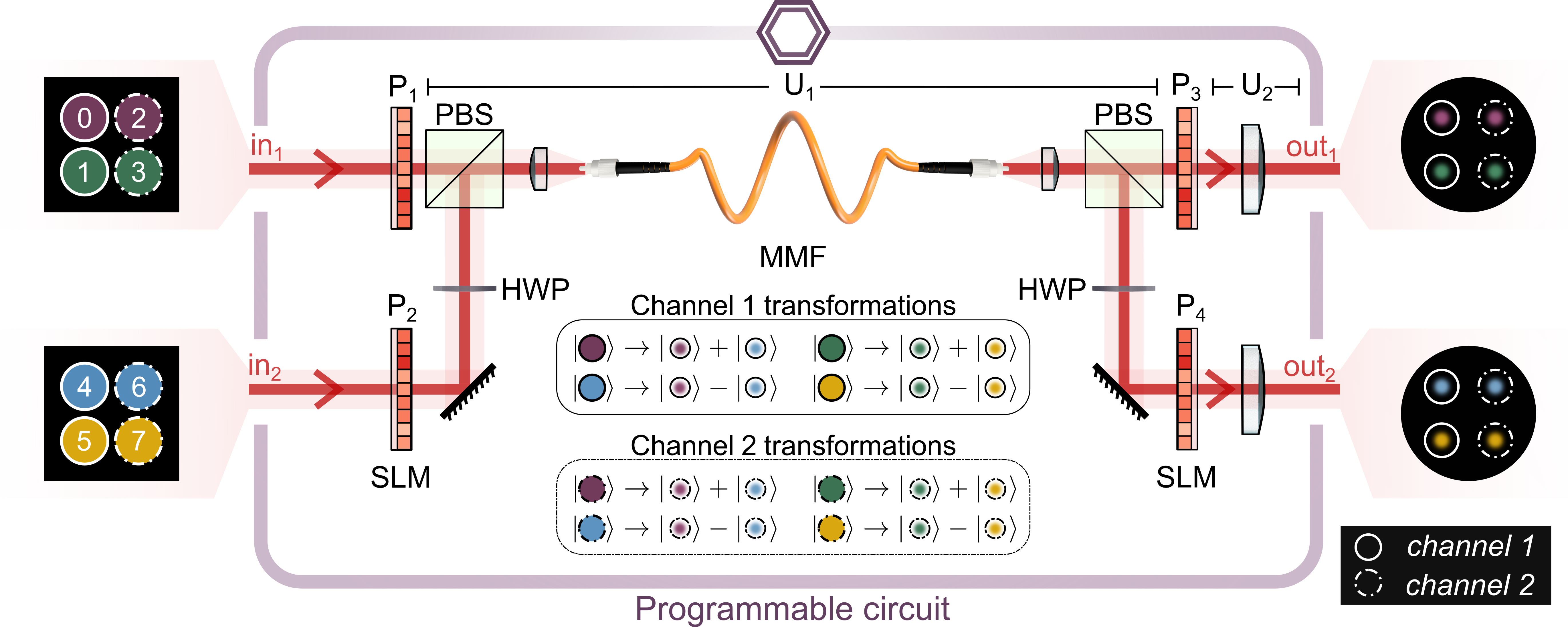}
    \caption{\textbf{Programmable multi-port circuit design}: A reconfigurable $8\times 8$-dimensional multi-port circuit is implemented in a top-down manner \cite{goel2022inverse} by placing a multi-mode fibre (MMF, $U_1$) between four programmable phase planes ($P_1$-$P_4$) displayed on spatial light modulators (SLMs). The circuit simultaneously operates on two input/output channels (1,2) each carrying 4-dimensional spatial modes (labelled 0 to 7). The circuit implements transformations of the form $\mathbb{T} = U_2(P_3+P_4)U_1(P_1+P_2)$ between input macro-pixel modes (left side) and focused Gaussian spots across the facet of the MMF (right side). The phase plane solutions ($P_1$-$P_4$) for a desired circuit transformation are obtained by using an inverse-design wavefront matching algorithm \cite{Kohtoku2005,Hashimoto2007}. The circuits programmed in this manner enable us to perform on-demand routing and interference operations on the 8 input/output spatial modes to realise the entanglement structures shown in Fig.~\ref{fig:NetStruct}c-f. Here we show the transformation corresponding to a simultaneous Bell-state measurement across the two channels, which implements the multiplexed entanglement swapping protocol shown in Fig.~\ref{fig:NetStruct}f. PBS: polarising beam splitter, HWP: half-wave plate.
    }
    \label{fig:ProgCirc}
\end{figure*}

Here we build on the top-down approach to implement a programmable $8\times 8$-dimensional multi-port device and use it to realise a prototype global multiplexed quantum network architecture that routes and swaps entanglement in a reconfigurable and multiplexed manner between two local networks of four users each. Our network architecture is depicted in Fig.~\ref{fig:NetStruct}a. Two local multi-user quantum networks are implemented via independent entanglement sources (S$_1$ and S$_2$) that distribute qubit-entanglement to four sets of users in each network (\{A$_1$, A$_2$, B$_1$, B$_2$\} and \{G$_1$, G$_2$, H$_1$, H$_2$\}).  A reconfigurable multi-port device (Fig.~\ref{fig:NetStruct}b) is implemented by placing an MMF (represented by the unitary operation U) between four programmable phase planes (P$_1$-P$_4$). The multi-port is used to connect the two local networks and realise a reconfigurable global network that can route and swap entanglement between all eight users in a multiplexed manner. The entanglement structures shown in Figs.~\ref{fig:NetStruct}c-f illustrate this versatile functionality, where a link between users (triangles) represents a shared qubit-entangled state. For example, multiplexed qubit entanglement can be shared between user pairs A$_1$B$_1$, A$_2$B$_2$, G$_1$H$_1$, and G$_2$H$_2$ (Fig.~\ref{fig:NetStruct}c). Alternatively, the network can be reconfigured to share qubit entanglement between user pairs A$_1$B$_1$, A$_2$G$_1$, B$_2$H$_1$, and G$_2$H$_2$ (Fig.~\ref{fig:NetStruct}e). Importantly, the multi-port device can also implement multiplexed entangling operations on independent input photons, enabling entanglement to be swapped between distant user pairs A$_1$H$_1$ and A$_2$H$_2$ (Fig.~\ref{fig:NetStruct}f).

This demonstration showcases the potential of complex-media-based platforms for realising large-scale and flexible quantum network architectures. Below, we describe the operation of our programmable multi-port device in detail and present results quantifying the performance of our reconfigurable global quantum network.

%%%%%%%%%%%%%%%%% CIRCUIT OPERATION %%%%%%%%%%%%%%
 \section{Programmable Multi-port Design}
We harness the complexity of the inter-modal coupling inside a random scattering medium to construct a top-down programmable device that can implement arbitrary high-dimensional quantum gates operating on two independent photons. A linear circuit can be represented by a transformation $\mathbb{T}$ that maps a set of input modes to a set of output modes. As depicted in Fig.~\ref{fig:ProgCirc}, our design consists of a 30cm-long multi-mode fibre (Thorlabs GIF625) acting as a large, ambient mode-mixer, placed between four phase planes implemented on programmable spatial light modulators (SLMs). The circuit's transformation can be decomposed as $\mathbb{T} = U_2(P_3+P_4)U_1(P_1+P_2)$, with $P_1-P_4$ representing the reconfigurable phase planes, $U_1$ describing the transmission through the MMF and associated coupling optics, and $U_2$ corresponding to the optical system between phase planes $P_3$/$P_4$ and the detection system.  

The circuit is designed by first characterising the MMF transmission matrix $U_1$ across all input and output modes. This is performed with classical illumination and a physics-based neural network approach that describes the optical system of the experiment~\cite{goel2023TM}. With complete knowledge of the scattering process in the MMF, desired target circuits $\mathbb{T}$ can be inverse-designed using iterative wavefront matching optimisations of input and output optical fields propagating forwards and backwards through the phase planes $P_1$-$P_4$~\cite{Kohtoku2005,Hashimoto2007}. The phase plane solutions obtained in this manner are implemented via holograms displayed on the programmable SLMs.

Our multi-port circuit implements high-dimensional quantum operations in the photonic transverse-spatial degree-of-freedom. Here, we are free to choose a suitable discretised spatial mode basis, which may be dictated by the type of multi-mode waveguides used in the network. At the input of the multi-port, we choose the macro-pixel basis~\cite{HerreraValencia2020} that is composed of localised circular modes (Fig.~\ref{fig:ProgCirc}, left). At the output of the multi-port (following a lens), we target modes randomly selected from a set of focused Gaussian spots distributed isometrically across the facet of the MMF (Fig.~\ref{fig:ProgCirc}, right). To use the full dimensionality of the MMF, we exploit the polarisation-dependence in the mode-mixing process of the MMF and associate the two polarisations to the two input ports ($in_1, in_2$), into which we inject two independent photons. At the output of the multi-port, these two polarisations also define the two output ports ($out_1, out_2$) used for directing photons to different users.

At each input port, we construct a four-dimensional spatial mode basis composed of macro-pixel modes $\{\ket{m}\}_{m}$, with $m=\{0,1,2,3\}$ labelling the modes at input 1, and $m=\{4,5,6,7\}$  labelling the modes at input 2. Each four-dimensional input set is split into two qubit subspaces, which are used as distinct channels for multiplexing qubit-entanglement in the network. As such, the multi-port operates on four distinct qubit channels composed of eight input macro-pixel modes. At input 1, we define the two channels as $\mbox{Ch}_{1} = \{\ket{0},\ket{1}\}$ and $\mbox{Ch}_{2} =\{\ket{2},\ket{3}\}$, while at input 2, we define $\mbox{Ch}_{1} = \{\ket{4},\ket{5}\}$ and $\mbox{Ch}_{2} =\{\ket{6},\ket{7}\}$ at input 2. The same is done with the eight foci modes at the output ports to define a total of four output qubit channels.

The multi-port circuit enables coherent control over the composite eight-dimensional modal space of two input photons. When these photons are obtained from two independent local networks, the multi-port acts as an interconnect, enabling multiplexed entanglement routing and swapping operations across the two local networks. Reconfigurable connectivity is implemented by programming different $8 \times 8$ unitary operations using the SLMs, without the need for any changes to the optical setup. For example, we can realise the entanglement structure shown in Fig.~\ref{fig:NetStruct}c with an Identity $\mathbb{T}_I$ gate. The entanglement structures shown in Fig.~\ref{fig:NetStruct}d-e are realised by programming the operations $\mathbb{T}_X$ and $\mathbb{T}_M$ (see Supplementary Information~\ref{sup_sec:exp} for operation definitions). As an example, here we define the operation $\mathbb{T}_S$ that allows us to perform the multiplexed entanglement swapping protocol between two pairs of distant users (Fig.~\ref{fig:NetStruct}f):
%To teleport entanglement and connect distant nodes A and H over two distinct channels (Fig.~\ref{fig:NetStruct}.f), we perform a multiplexed entanglement swapping protocol by targeting the following operation:

\begin{equation}
\label{eq:swap_circuit_defn}
\mathbb{T}_S = \; 
\frac{1}{\sqrt{2}}
    \begin{bmatrix}
        1 & 0 & 0 & 0 & 1 & 0 & 0 & 0\\
        0 & 1 & 0 & 0 & 0 & 1 & 0 & 0\\
        0 & 0 & 1 & 0 & 0 & 0 & 1 & 0 \\
        0 & 0 & 0 & 1 & 0 & 0 & 0 & 1 \\
        1 & 0 & 0 & 0 & -1 & 0 & 0 & 0\\
        0 & 1 & 0 & 0 & 0 & -1 & 0 & 0\\
        0 & 0 & 1 & 0 & 0 & 0 & -1 & 0 \\
        0 & 0 & 0 & 1 & 0 & 0 & 0 & -1 \\
    \end{bmatrix}.
\end{equation}

The operation of this $8\times 8$ circuit is illustrated in Fig.~\ref{fig:ProgCirc}. As can be seen, this circuit is implementing two simultaneous beam-splitter operations across the two input/output channels, mixing the eight input macro-pixel modes unitarily with each other. This serves as a multiplexed entangling gate, enabling parallel Bell-state measurements to be made on each channel. More details on the multiplexed entanglement swapping protocol can be found in Section~\ref{sup_sec:swap}. 

%%%%%%%%%%%%%%%%%%%%%%%%% Networking results%%%%%%%%%%%%%%

\section{Global Reconfigurable Entanglement Network}\label{sec:ent_distribution}

We demonstrate the operation of the programmable multi-port circuit by connecting two local, four-user multiplexed entanglement networks to realise a global multiplexed network of eight users (Fig.~\ref{fig:NetStruct}a) with several different network configurations (Fig.~\ref{fig:NetStruct}c-f). The local networks are implemented by generating high-dimensional ($d=4$) spatial-mode entanglement in two nonlinear crystals (S$_1$ and S$_2$) via the process of spontaneous parametric down-conversion (SPDC). Each $d=4$ bipartite entangled state is multiplexed into two qubit-entanglement channels and distributed to the four users in each local network. The users in local networks 1 and 2 are labelled as  \{A$_1$, A$_2$, B$_1$, B$_2$\} and \{G$_1$, G$_2$, H$_1$, H$_2$\}. Each user is equipped with a detection system for performing projective measurements of arbitrary spatial modes using an SLM, single-mode fibre, and single-photon detector. For a detailed description of our entanglement generation and measurement methods, please see~\ref{sup_sec:exp}.

\begin{figure*}[ht!]
    \centering
    \includegraphics[width=0.8\linewidth]{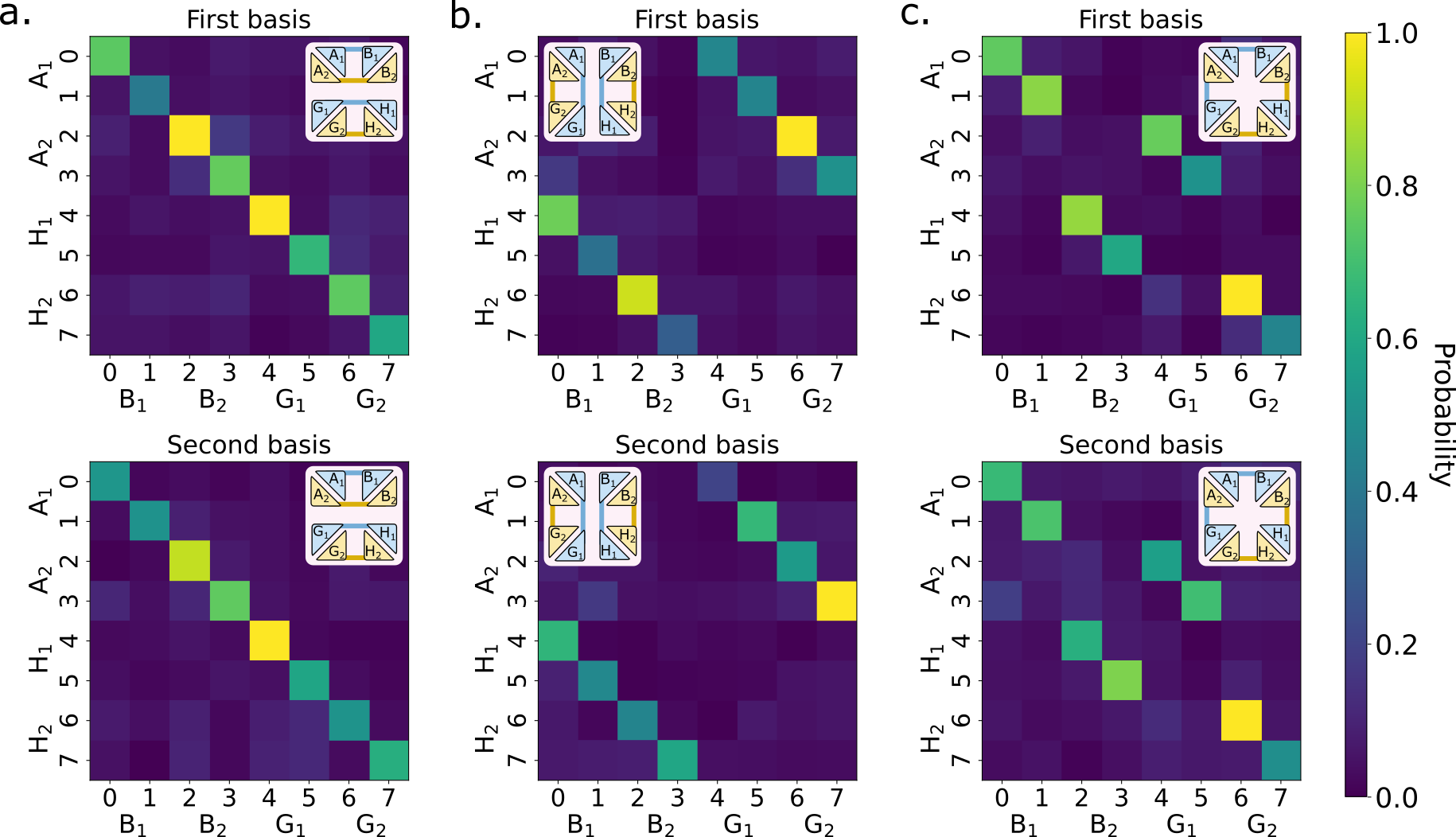}
    \caption{\textbf{Multiplexed entanglement routing}. We program $8\times 8$-dimensional circuit operations that route qubit entanglement between eight users in three different global network configurations (insets). These include: a) entanglement links between four local user pairs AB and GH, b) four global user pairs AG and BH, and c) four pairs consisting of both local and global user combinations. Normalised two-photon correlations measured in two mutually unbiased bases (MUBs) are shown for each network configuration. By applying a suitable entanglement witness, we are able to certify the successful routing of entanglement in all three configurations.}
    \label{fig:EntDist}
\end{figure*}

The programmable multi-port operates on two independent photons distributed from S$_1$ to users B$_1$, B$_2$ and S$_2$ to users G$_1$, G$_2$. By implementing the circuits $\mathbb{T}_I$, $\mathbb{T}_X$, and $\mathbb{T}_M$, we realise the three multiplexed entanglement structures shown in Fig.~\ref{fig:NetStruct}c-e. We quantify the quality of the network structures by measuring two-photon correlations between the eight users for each chosen multiplexed network configuration and applying suitable entanglement witnesses. The implementation of the first network structure via circuit operation $\mathbb{T}_I$ is illustrated in Fig.~\ref{fig:EntDist}a, where measurements in two mutually unbiased bases (MUBs) show correlations between local user pairs A$_1$B$_1$, A$_2$B$_2$, G$_1$H$_1$, and G$_2$H$_2$. The network is then reconfigured to realise the first global network structure via circuit operation $\mathbb{T}_X$, with strong two-photon correlations observed between user pairs from different local networks: A$_1$G$_1$, A$_2$G$_2$, B$_1$H$_1$, and B$_2$H$_2$ (Fig.~\ref{fig:EntDist}b). Finally, the global network structure corresponding to multiplexed entanglement links across both local and global networks is realised via the circuit operation $\mathbb{T}_M$, with two-photon correlations observed between user pairs A$_1$B$_1$, A$_2$G$_1$, B$_2$H$_1$, and G$_2$H$_2$ (Fig.~\ref{fig:EntDist}c).

\begin{smallboxtable}{Fidelities [\%] to the maximally entangled state of the states shared in the multiplexed programmable network using $8$-dimensional gates}{F8x8gates}
\begin{centering}
\begin{tabular}{|c|c|c|c|c|}
\hline
\multirow{2}{*}{$\mathbb{T}_I$} & A$_1$B$_1$                       & A$_2$B$_2$     & G$_1$H$_1$       & G$_2$H$_2$       \\ \cline{2-5} 
                              & 86.0 $\pm$ 1.0 & 78.6 $\pm$ 0.9 & 86.3 $\pm$ 0.9 & 79.2 $\pm$ 1.1 \\ \hline
\multirow{2}{*}{$\mathbb{T}_X$} & A$_1$G$_1$                       & A$_2$G$_2$       & B$_1$H$_1$     & B$_2$H$_2$      \\ \cline{2-5} 
                              & 81.6 $\pm$ 1.0                  & 79.2 $\pm$ 1.1 & 81.7 $\pm$ 1.2 & 80.8 $\pm$ 1.1 \\ \hline
\multirow{2}{*}{$\mathbb{T}_M$} & A$_1$B$_1$                       & A$_2$G$_1$       & B$_2$H$_1$       & G$_2$H$_2$      \\ \cline{2-5} 
                              & 82.3 $\pm$ 1.0                  & 81.2 $\pm$ 1.2 & 76.5 $\pm$ 1.2 & 81.4 $\pm$ 1.0 \\ \hline
 \multirow{2}{*}{$\mathbb{T}_S$} & \multicolumn{2}{c|}{A$_1$H$_1$} &  \multicolumn{2}{c|}{A$_2$H$_2$} \\ \cline{2-5}
 &\multicolumn{2}{c|}{77.1 $\pm$ 3.3} & \multicolumn{2}{c|}{83.2 $\pm$ 2.7} \\
 \hline 
\end{tabular}
\end{centering}
\vspace*{5pt}
\footnotesize{\\ *Errors are reported to one standard deviation}
\end{smallboxtable}

While Fig.~\ref{fig:EntDist} shows correlations in two MUBs, we also measure correlations in the third MUB of each two-dimensional subspace. This allows us to calculate the exact fidelity to a target entangled state via a fidelity witness~\cite{Bavaresco:2018gw} and certify the quality of the multiplexed entanglement distribution in all network configurations. As shown in Table.~\ref{table:F8x8gates}, we estimate the fidelities of the entangled states shared between all user pairs to the two-dimensional maximally entangled state. For circuit operation $\mathbb{T}_I$, we obtain entangled state fidelities of 78.6\% or higher, demonstrating the successful distribution of multiplexed qubit-entanglement between all four local user pairs. The first global network configuration realised by circuit operation $\mathbb{T}_X$ results in entangled state fidelities of 79.2\% or higher, demonstrating that switching from a local to a global structure does not compromise the entanglement quality. Finally, the local-global configuration implemented by circuit operation $\mathbb{T}_M$ also results in qubit-entanglement being certified with a fidelity of 76.5\% or higher, demonstrating the versatility of the multi-port circuit. We also test the ability of the multi-port circuit to route qubit and qutrit ($d=3$) entanglement over a single channel (without multiplexing) in two different network configurations of four users. As shown in Supplementary section ~\ref{sup_sec:routing}, we able to certify successful routing of qubit and qutrit entanglement for both network configurations. Errors in the fidelities are calculated using a Monte Carlo simulation that takes into account Poissonian photon-counting statistics.

\begin{figure*}[ht!]
    \centering
    \includegraphics[width=0.8\linewidth]{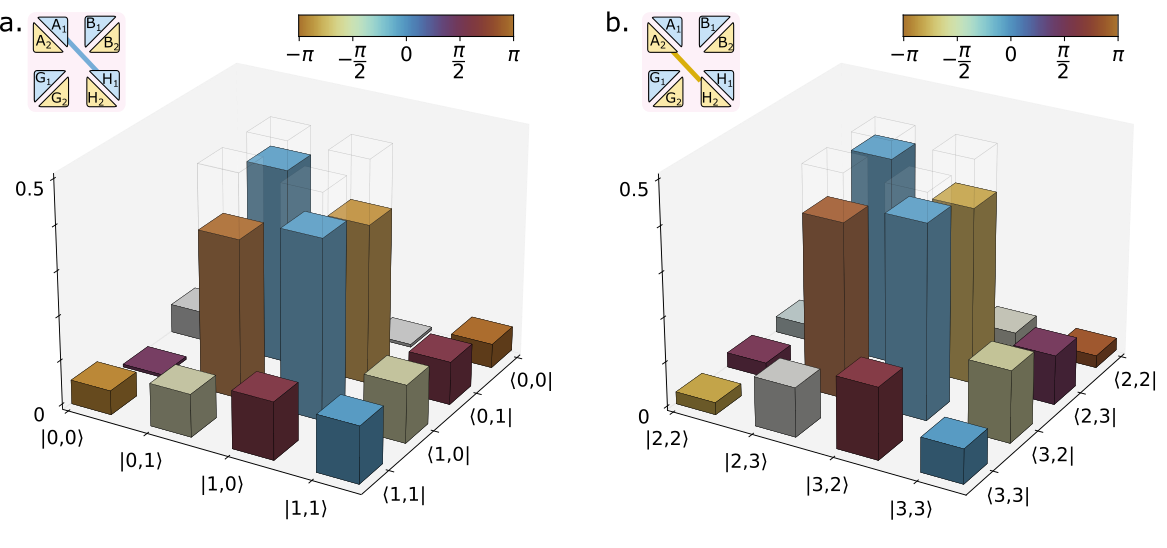}
    \caption{\textbf{Multiplexed entanglement swapping}: Reconstructed density matrices of the qubit-entangled states shared between distant user pairs a) A$_1$H$_1$ and b) A$_2$H$_2$, via the circuit operation implementing simultaneous Bell-state measurements on two multiplexed channels. We obtain fidelities to the two-dimensional maximally entangled target states $|\Psi_T\rangle$ (transparent bars) of 77.1$\pm$ 9.8\% for channel 1 and 83.2$\pm$ 8.3\% for channel 2, demonstrating the success of our entanglement swapping protocol.}
    \label{fig:Swap}
\end{figure*}

%%%%%%% Swapping on a 4-dim space %%%%%%%%%
Finally, we implement the circuit operation $\mathbb{T}_S$ to realise multiplexed entanglement swapping between distant user pairs A$_1$H$_1$ and A$_2$H$_2$, connecting the edges of the two local networks with entanglement links. The swapping operation is conditioned on successful multiplexed Bell-state measurements between user pairs B$_1$G$_1$ and B$_2$G$_2$. To certify the quality of this challenging network configuration, we perform two sets of four-photon correlation measurements across the two network channels. This allows us to characterise the swapped entangled states between the distant users via quantum state tomography (see details in~\ref{sup_sec:QST}). Fig~\ref{fig:Swap} shows the reconstructed density matrices of the two entangled states shared between A$_1$H$_1$ and A$_2$H$_2$. The estimated states in each channel have fidelities to a maximally entangled target state of the form $|\Psi_T\rangle = \frac{1}{\sqrt{2}}\left(\ket{0_A1_H} - e^{i\theta}\ket{1_A0_H} \right)$ of more than 77\% (with the phase $\theta = 1.535\pi$ arising from the relative phase between input polarisations). This successfully demonstrates that our multi-port circuit is able to perform multiplexed entangling operations that are critical for interconnecting several independent local quantum networks.

Additionally, we test the stability of the programmable circuit over time (see \ref{sup_sec:stability}) by repeatedly implementing all $8\times 8$ bipartite routing operations and measuring the fidelity to the maximally entangled state of all distributed entangled states over a period of 14 days. During this time, the mechanical stability of the MMF is maintained with simple coupling stages or clamps to the optical table. We observe that the fidelity of the distributed entangled states remains well above the bound, without the need for re-alignment or re-characterisation of the MMF transmission matrix (U), demonstrating the robustness of our circuit implementation. 

\section{Conclusion and Outlook}

We have demonstrated a reconfigurable multiplexed photonic network that connects two local four-user networks to realise a global eight-user network, allowing distant pairs of users to share entanglement in a multiplexed and flexible manner. Our network demonstration employs a unique multi-port device that harnesses the large mode-mixing process inside a multi-mode fibre to implement high-dimensional operations on two independent photons carrying eight spatial modes. We achieve multiplexed routing and swapping of qubit entanglement with fidelities to the maximally entangled state of above 76\% for all network configurations and channels. We also demonstrate the routing of qutrit entanglement, showcasing the versatility our MMF-based device. To enable multiplexed operation, two-dimensional spatial-mode subspaces are used as distinct entanglement distribution channels, allowing users to share multiple qubit-entangled states simultaneously. Moreover, our ability to perform programmable Bell-state measurements in parallel enables us to link four remote users through simultaneous entanglement swapping over two independent channels. 

Our multi-port device allows for efficient integration into existing communications infrastructure because of its straightforward coupling to optical fibre links. The stability of our system is maintained over weeks, and there is no need for special protective enclosures or re-calibration of the mode-mixing process. With our top-down approach, the fidelity and size of the operations can be improved with a higher number of programmable phase layers within the circuit or a larger dimensionality in the auxiliary modal space offered by the MMF \cite{goel2022inverse}. However, effects such as loss or modal dispersion become more prevalent in these regimes, and shorter fibre lengths or spectral-temporal control techniques may be necessary. Nevertheless, our device serves as a powerful approach towards realising large-scale, multi-user entanglement networks with reconfigurable connectivity. As the number of modes and users increase, the characterisation of such a device may require efficient process fidelity witnesses based on minimal measurements \cite{Engineer_2024}. Future implementations of such a multi-port device could also explore entanglement in different high-dimensional spatial-mode bases \cite{Malik_Boyd_2014,PhysRevA.108.052612}, and other photonic degrees-of-freedom such as time-energy \cite{Jha_2008}. Our implementation demonstrates the potential of complex-media-based platforms for realising practical and noise-robust quantum network architectures \cite{Srivastav_2022} and the control and distribution of large quantum photonic states.

\textit{Acknowledgements:}
We would like to thank Sophie-Elisabeth Lerchbaumer for her help with the early stages of the experiment. We acknowledge financial support from the UK Engineering and Physical Sciences Research Council (EPSRC) (EP/P024114/1), European Research Council (ERC) Starting Grant PIQUaNT (950402), and the Royal Academy of Engineering Chair in Emerging Technologies programme (CiET-2223-112).

\bibliographystyle{apsrev4-1fixed_with_article_titles_full_names}
\bibliography{MPQPN}

%%%%%%%%%%%%%%%%%%%%% APPENDIX %%%%%%%%%%%%%%%%%%%%%%%%%%%%%%%%
\clearpage
\onecolumngrid
\appendix
\input{MPQPN_supp}

\end{document}

%% file: MPQPN_supp.tex
\renewcommand{\appendixpagename}{\begin{center}\large{\textbf{Supplementary information for: A Multiplexed Programmable Quantum Photonic Network}} \end{center}\label{SI}}

\appendixpagename

\renewcommand{\thesubsection}{S.\arabic{section}.\arabic{subsection}}
\renewcommand{\thesection}{S.\arabic{section}}

\setcounter{equation}{0}
\setcounter{figure}{0}
\setcounter{table}{0}
\renewcommand{\theequation}{S.\arabic{section}.\arabic{equation}}
\renewcommand{\thetable}{S.\arabic{table}}
\renewcommand{\thefigure}{S.\arabic{figure}}
\renewcommand{\theHfigure}{S.\arabic{figure}}

The following supplementary information is provided: Details of the experimental setup~(\ref{sup_sec:exp}), techniques for characterising the complex scattering in the multi-mode fibre and designing optical circuits~(\ref{sup_sec:TMandWFM}), results for two and three-dimensional entanglement routing over a single channel~(\ref{sup_sec:routing}), performance of the circuit over time~(\ref{sup_sec:stability}), single and two-channel entanglement swapping~(\ref{sup_sec:swap}), and method for estimating the swapped states through quantum state tomography~(\ref{sup_sec:QST}).

\section{Detailed experimental setup}
\label{sup_sec:exp}

\begin{figure*}[ht!]
    \centering
    \includegraphics[width=\textwidth]{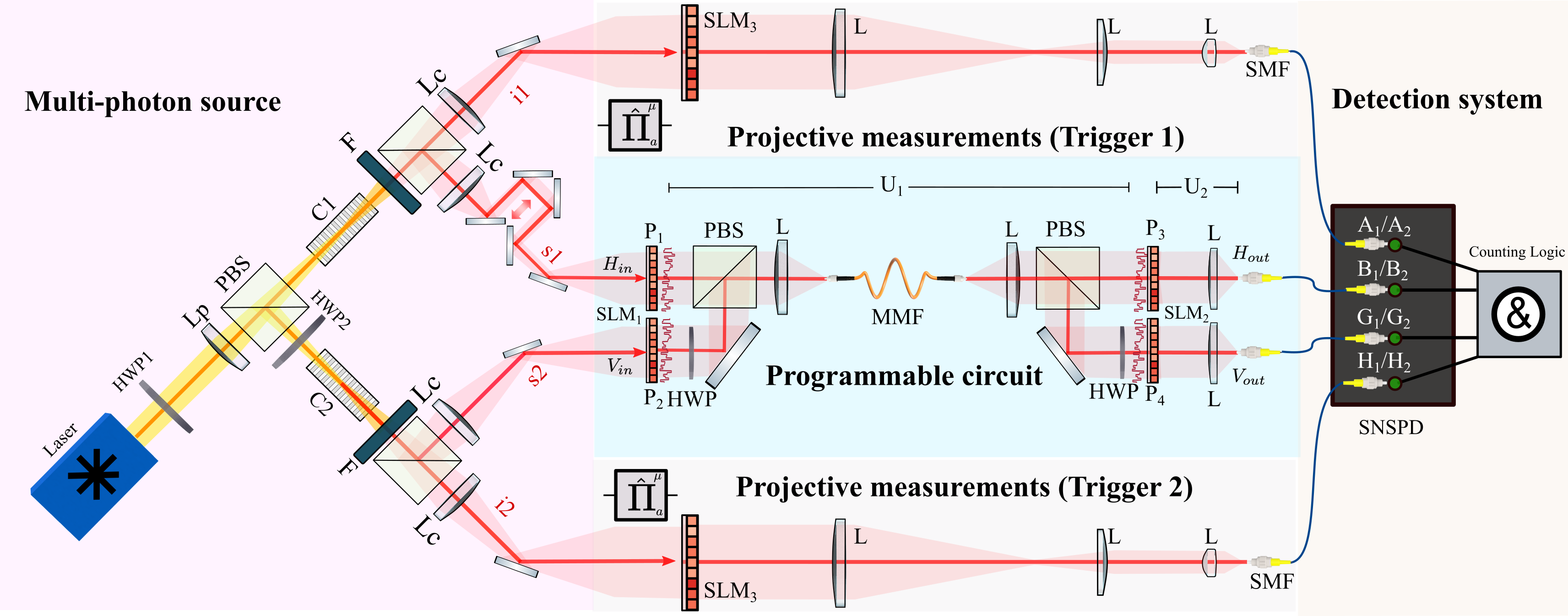}
    \caption{\textbf{Experimental implementation} Multi-user local networks are implemented by generating high-dimensional spatial-mode entanglement through spontaneous parametric down-conversion (SPDC) in periodically poled Potassium Titanyl Phosphate (ppKTP) crystals (C1 and C2). The signal photons from each pair ($s_1$ and $s_2$) are sent to users \{B$_1$,B$_2$,G$_1$,G$_2$\} through the programmable circuit composed of a multi-mode optical fibre (MMF) and two spatial light modulators divided into two screens each: SLM$_1$:P$_{1,2}$ and SLM$_2$:P$_{3,4}$). To ensure both input photons are indistinguishable through the circuit, identical paths are ensured with a delay stage on the path of $s_1$. Users \{B$_1$,B$_2$,G$_1$,G$_2$\} use phase screens P$_3$ and P$_4$ to project over different output modes. The other two photons from each entangled state ($i_1$ and $i_2$) are sent directly to users \{A$_1$,A$_2$,H$_1$,H$_2$\}, where projective measurements in spatial modes are performed by combining a spatial light modulator (SLM$_3$) and single-mode fibres (SMF). Note that SLM$_3$ is divided into two sections, each for manipulating a different idler photon. Photons arriving at each user pair are coupled to SMFs that guide them to superconducting nanowire single-photon detectors (SNSPD). Coincidence events between the eight users are registered through a time-tagging counting logic. L: lens, PBS: polarised beam-splitter, F: filter}
    \label{fig:ExpSetup}
\end{figure*}

The experimental setup for our programmable network is illustrated in~\ref{fig:ExpSetup}. The setup can be divided into three sections: 

\begin{itemize}
    \item \textbf{Entangled photon sources:} A four-photon state composed of two independent pairs of high-dimensional entangled photons generated through Type II SPDC at 1550~nm in two identical non-linear periodically poled Potassium Titanyl Phosphate (ppKTP) crystals (1~mm$\times$2~mm$\times$5~mm).
    
    \item \textbf{Programmable circuit:} Using a top-down, inverse-design approach \cite{goel2022inverse}, a 30~cm-long graded-index fibre (Thorlabs GIF625) acting as a complex mode-mixer, and supporting approximately 150 modes per polarisation at 1550~nm, is placed between two programmable phase layers comprising of two phase planes each ($\{P_1, P_2\}$ and $\{P_3, P_4\}$) implemented on two spatial-light-modulators (Hamamatsu X10468-08).
    
    \item \textbf{Detection system:} Each of the eight users can perform single-outcome spatial-mode projections with the combination of programmable phase-screens implemented on spatial light modulators (SLMs), single-mode fibres (SMFs), and superconducting nanowire detectors (SNSPDs) \cite{Bouchard_2018}. Pairs of users \{(A$_1$,A$_2$), (B$_1$,B$_2$), (G$_1$,G$_2$), (H$_1$,H$_2$)\} use one phase-screen each to perform projective measurements in their respective two-dimensional subspace. For the classical characterisation of the complex mode-mixing in the fibre and initial testing of the gates, a removable mirror is used to propagate laser light through the circuit, with an InGaAs camera (Allied Vision, Goldeye G-008 Cool TEC1) at the output.  
\end{itemize}

\subsection{Multi-photon source}
A 775-nm Ti:Sapphire femtosecond pulsed laser (140fs) is divided into two beams with a polarised beam-splitter (PBS) and a half-wave plate (HWP1). Each beam is focused with a power of 750~mW onto each ppKTP crystal (1mm $\times$ 2mm $\times$ 5mm) to generate independent pairs of entangled photons. With a lens L$_p$ ($f_p =$150~mm) we set the beam waist ($1/e^2$ radius) of the pump at the centre of each crystal to $w_p = 62\mu$m. With this waist, we estimate a Schmidt number of each of the generated biphoton states of $K_G\sim 6$~\cite{Srivastav2021,Law2004}. Since the collection optics will further reduce the available entanglement, this modal bandwidth is an upper bound of the dimensionality we expect from each entangled pair. 

The energy- and phase-matching conditions for these pump and crystal parameters lead to some non-separability in the joint-spectral amplitude (JSA), thus reducing heralded single-photon purity. This limits the Hong-Ou-Mandel interference visibility and the subsequent entanglement swapping fidelity. This reasoning is more nuanced when considering multiple spatial modes, whose JSAs may be distinct. However, our configuration minimises the spatial-spectral coupling across the spatial modes of interest~\cite{Srivastav2021} so the JSAs should be well approximated by the collinear JSA. A natural route to improving this purity would be to reduce the spectral bandwidth of the pump, decrease the crystal length, or introduce spectral bandpass filters on the photon(s)~\cite{Graffitti2018}. Domain-engineered crystals could also be employed to tailor the crystal nonlinearity, eliminating the need for narrowband spectral filtering. This would reduce losses while improving the overall signal-to-noise ratio~\cite{Graffitti:2018ib}

An alternative route to spectral filtering originates from the finite spectral bandwidth of the scattering medium\cite{Andreoli2015,Carpenter2016}. Since the operations programmed in the MMF are optimised for the central wavelength at which it was classically characterised, the reduced mode transformation efficiency at more distant frequencies leads to effective spectral filtering into the desired output mode. This effect was sufficient to observe approximately 10~nm Gaussian spectral filtering on our programmed identity gate, allowing us to obtain high-quality swapped state fidelities even after the removal of spectral bandpass filters used for the classical characterisation. The precise nature of these effects depends intimately on the spatial modes in use, gate transformation implemented, the dispersive properties of the random configuration of the MMF, as well as the spectral-spatial properties of the bi-photon state generated in the crystal, as such, the contributions to our swapped state fidelities cannot be exactly isolated.    

After the crystal, the pump is dumped using a dichroic mirror (corresponding to filter F in the figure), and the signal and idler photons of each source are separated using a PBS. The signal photons $s_1$ and $s_2$ are injected into the optical circuit with orthogonal polarisations (more details in the following section). To control their indistinguishability, we introduced a temporal delay with a motorised stage on the path of $s_1$. The idler photons $i_1$ and $i_2$ are manipulated with an SLM and SMF combination, allowing for single outcome projective measurements in arbitrary spatial modes. Taking into account the physical parameters of the SPDC generation, we estimate the correlation bandwidth $\sigma_S$ of the biphoton states at the Fourier plane and use a set of three lenses between crystal and SLMs to adequately collect and resize the pair of photons such that they cover most of the area of each SLM screen ($600 \times 400$ pixels with 20~$\mu$m pixel pitch), but avoids any clipping. For the sake of simplicity, this optical system is represented in~\ref{fig:ExpSetup} by the lens $L_c$. After reflection from SLMs, a telescope system and an aspheric lens are used for mode-matching the projected photons to either SMF or MMF collection modes (In~\ref{fig:ExpSetup}, this optical system is labeled as $L$ inside the programmable circuit section, and shown explicitly with the three lenses in the projective measurements section). 

\subsection{Programmable optical circuit}

A top-down programmable optical circuit~\cite{goel2022inverse} harnesses random scattering inside a multi-mode fibre. The circuit is designed to operate on the macro-pixel basis on the input side, while the target output modes are randomly selected from the set of all possible focused spots (foci) distributed isometrically across the facets of the MMF. Note that our fibre only supports approximately 150 spatial modes per polarisation, but we over-sample the facet of the fibre to improve the characterisation of the transmission matrix.

The mode-mixing process inside the fibre is dependent on polarisation. Light with horizontal polarisation has a different transmission matrix than light with vertical polarisation. Furthermore, modes with a given polarisation will exit the fibre in two different polarisations because polarisation isn't preserved. This polarisation mixing allows us to take advantage of the full modal space offered by the multi-mode fibre by defining the ports of the circuit according to polarisation~\cite{Xiong2018,Leedumrongwatthanakun:2019wt}. As shown in~\ref{fig:ExpSetup}, we have two input photon ports (Input 1 is $H_{in}$ and Input 2 $V_{in}$) and two output photon ports (Output 1 is $H_{out}$, and Output 2 is  $V_{out}$).  

The screen of SLM$_1$ is divided into two regions to be used as phase planes $P_1$ and $P_2$, allowing for the independent control of light entering from each input port. While the signal photons ($s_1$ and $s_2$) are generated with the same polarisation in their respective crystal, we rotate the polarisation of one of the photons (not shown in the figure), superposed them with a polarised beam splitter (PBS), and inject them into the MMF with orthogonal polarisations. Light coming out of the fibre is separated using a second PBS, with each output polarisation reflecting on a different region of SLM$_2$ to be used as phase planes $P_3$ and $P_4$, and directed to its corresponding detection stage.  

We implement four different kinds of circuits defined by operations $\{\mathbb{T}_I,\mathbb{T}_X,\mathbb{T}_M,\mathbb{T}_S\}$, with:
\begin{equation}
    \label{eq:circuit_defn_supp}
\mathbb{T}_I = \;
    \begin{bmatrix}
        1 & 0 & 0 & 0 & 0 & 0 & 0 & 0\\
        0 & 1 & 0 & 0 & 0 & 0 & 0 & 0\\
        0 & 0 & 1 & 0 & 0 & 0 & 0 & 0 \\
        0 & 0 & 0 & 1 & 0 & 0 & 0 & 0 \\
        0 & 0 & 0 & 0 & 1 & 0 & 0 & 0\\
        0 & 0 & 0 & 0 & 0 & 1 & 0 & 0\\
        0 & 0 & 0 & 0 & 0 & 0 & 1 & 0 \\
        0 & 0 & 0 & 0 & 0 & 0 & 0 & 1 \\
    \end{bmatrix},
 \quad  \quad
\mathbb{T}_X = \;
\frac{1}{\sqrt{2}}
    \begin{bmatrix}
        0 & 0 & 0 & 0 & 1 & 0 & 0 & 0\\
        0 & 0 & 0 & 0 & 0 & 1 & 0 & 0\\
        0 & 0 & 0 & 0 & 0 & 0 & 1 & 0 \\
        0 & 0 & 0 & 0 & 0 & 0 & 0 & 1 \\
        1 & 0 & 0 & 0 & 0 & 0 & 0 & 0\\
        0 & 1 & 0 & 0 & 0 & 0 & 0 & 0\\
        0 & 0 & 1 & 0 & 0 & 0 & 0 & 0 \\
        0 & 0 & 0 & 1 & 0 & 0 & 0 & 0 \\
    \end{bmatrix},
 \quad  \quad 
 \mathbb{T}_M = \;
 \frac{1}{\sqrt{2}}
    \begin{bmatrix}
        1 & 0 & 0 & 0 & 0 & 0 & 0 & 0\\
        0 & 1 & 0 & 0 & 0 & 0 & 0 & 0\\
        0 & 0 & 0 & 0 & 1 & 0 & 0 & 0 \\
        0 & 0 & 0 & 0 & 0 & 1 & 0 & 0 \\
        0 & 0 & 1 & 0 & 0 & 0 & 0 & 0\\
        0 & 0 & 0 & 1 & 0 & 0 & 0 & 0\\
        0 & 0 & 0 & 0 & 0 & 0 & 1 & 0 \\
        0 & 0 & 0 & 0 & 0 & 0 & 0 & 1 \\
    \end{bmatrix},
\end{equation}

The operation for the multiplexed entanglement swapping protocol  $\mathbb{T}_S$ is defined in Eq.\ref{eq:swap_circuit_defn} of the main text. Labels indicating the ordering of modes and channels are shown in Table~\ref{table:swapgate8dim}.

\begin{table}[ht!]
\begin{tabular}{|ccc|cccccccc|}
\hline
                                            &                                           &       & \multicolumn{4}{c|}{In1}                                & \multicolumn{4}{c|}{In2}                               \\ \cline{2-11} 
\multicolumn{1}{|c|}{}                      & Channels                                  &       & \multicolumn{2}{c|}{Ch 1}   & \multicolumn{2}{c|}{Ch 2}   & \multicolumn{2}{c|}{Ch 1}    & \multicolumn{2}{c|}{Ch 2} \\ \cline{3-11} 
\multicolumn{1}{|c|}{}                      & \multicolumn{1}{c|}{}                     & Modes & 0 & \multicolumn{1}{c|}{1} & 2 & \multicolumn{1}{c|}{3} & 4  & \multicolumn{1}{c|}{5} & 6           & 7          \\ \hline
\multicolumn{1}{|c|}{\multirow{4}{*}{\rotatebox[origin=c]{90}{Out 1}}} & \multicolumn{1}{c|}{\multirow{2}{*}{Ch 1}} & 0     & 1 & 0                      & 0 & 0                      & 1  & 0                      & 0           & 0          \\
\multicolumn{1}{|c|}{}                      & \multicolumn{1}{c|}{}                     & 1     & 0 & 1                      & 0 & 0                      & 0  & 1                      & 0           & 0          \\ \cline{2-3}
\multicolumn{1}{|c|}{}                      & \multicolumn{1}{c|}{\multirow{2}{*}{Ch 2}} & 2     & 0 & 0                      & 1 & 0                      & 0  & 0                      & 1           & 0          \\
\multicolumn{1}{|c|}{}                      & \multicolumn{1}{c|}{}                     & 3     & 0 & 0                      & 0 & 1                      & 0  & 0                      & 0           & 1          \\ \cline{1-3}
\multicolumn{1}{|c|}{\multirow{4}{*}{\rotatebox[origin=c]{90}{Out 2}}} & \multicolumn{1}{c|}{\multirow{2}{*}{Ch 1}} & 4     & 1 & 0                      & 0 & 0                      & -1 & 0                      & 0           & 0          \\
\multicolumn{1}{|c|}{}                      & \multicolumn{1}{c|}{}                     & 5     & 0 & 1                      & 0 & 0                      & 0  & -1                     & 0           & 0          \\ \cline{2-3}
\multicolumn{1}{|c|}{}                      & \multicolumn{1}{c|}{\multirow{2}{*}{Ch 2}} & 6     & 0 & 0                      & 1 & 0                      & 0  & 0                      & -1          & 0          \\
\multicolumn{1}{|c|}{}                      & \multicolumn{1}{c|}{}                     & 7     & 0 & 0                      & 0 & 1                      & 0  & 0                      & 0           & -1         \\ \hline
\end{tabular}
\caption{Transformation between input and output modes over two channels implemented with operation $\mathbb{T}_S$.}
\label{table:swapgate8dim}
\end{table}

\subsection{User node: Detection and state analysis}
Projective spatial measurements on idler photons. $\hat{\Pi}_a^\mu$ in any spatial mode $a$ of basis $\mu$, are performed using SLM$_3$ and coupling to SMFs, which guide the photons to users $A_{1,2}$ and $H_{1,2}$. For the pair of photons going through the circuit, phase planes $P_1$-$P_4$ are used to display the wavefront matching solutions of the targeted operation, while simultaneously projecting the light from the two selected output ports to couple to SMFs that are guided towards users $B_{1,2}$ and $G_{1,2}$. Photons distributed to each user are sent to superconducting nanowire single-photon detectors (Quantum Opus, Opus One, efficiency >85\% at 1550~nm). A counting logic (Swabian, TimeTagger Ultra) records coincidences between the 8 users with a window of 200~ps. 

We note that due to our users' detection devices containing only one SNSPD per node, we perform the tomography/witnesses for each state/channel separately, whereas equipping each user with detectors would enable their multiplexed states to be recorded simultaneously/in parallel, with no reconfiguring of the programmable circuit.

When sending light from a classical source through the system, a removable mirror allows us to switch, without introducing changes in the circuit, from single-photon detection to imaging of the output speckle patterns using an InGaAs camera. A 400~mm lens focuses light reflecting from phase planes $P_{3,4}$ onto the sensor of the camera, where images of the output facet of the fibre are taken for the characterisation of $U_1$ and the initial test of the circuit's functionality.

\section{Characterization of complex mode-mixer and inverse design of optical circuit}
\label{sup_sec:TMandWFM}

To characterise the transformation $U_1$ across all input spatial and polarisation modes, we change the multi-photon source to a laser source ($\lambda = 1550 \pm 3 nm$) that maintains the paths shown in ~\ref{fig:ExpSetup} for $s_1,s_2$. We then allow the light from $s_i$ to be injected into the MMF, one by one for $i=1,2$ while random patterns $x_{1_i}$, $x_{2_i}$ are displayed on phase-planes $P_1/P_2$ and $P_3/P_4$ respectively.
The modulated light at the output is measured on the camera.
We employ a multi-plane neural network (MPNN)~\cite{goel2023TM} to learn the transmission matrix of the MMF for each input polarization $U_{1_i}$ by optimizing the following cost function
\begin{equation}
    \label{eq:mpnn_opt}
    \underset{U_{1_i}}{\text{min}}\quad
    \left|y_i-|U_2(x_{2_i}\odot (U_{1_i} x_{1_i})) |^2\right|
\end{equation}
Stacking the transmission matrices from each polarization $U_{1_i}$ allows us to construct $U_1$. This method enables us to bypass the need for an external reference with a machine-learning model that describes the optical system of the experiments. 

Since the transmission matrices $U_{1_i}$ are measured independently, we have no information on the relative phase between different input polarisations. While this phase doesn't affect the performance of the circuit when routing or switching entanglement, the effect on the entanglement swapping protocol is to introduce a phase on the state between Ada and Hedy of the form $\ket{\Psi} = \left(\ket{01} + e^{i\varphi} \ket{10}\right)$. 

To make sure the classical characterisation of the spatial mode-mixing process inside the fibre reproduces the behaviour of the SPDC photons, we use a system of telescopes for careful mode-matching between the laser source and the estimated spatial distribution of the multimodal SPDC emission~\cite{Srivastav2021}. To characterise the coupling loss in the circuit, we measure the coupling efficiency from the free-space single-mode laser source to the MMF to be above 76\%. 

With the complete knowledge of the scattering processes in the mode-mixer, target circuits can be inverse-designed using phase pattern solutions for planes $P_1, P_2, P_3$ and $P_4$ that are calculated with iterative wavefront matching~(WFM) optimisations of input and output optical fields~\cite{goel2022inverse}. For this, input and output modes are generated for a chosen set of discrete spatial mode bases and target transformation $\mathbb{T}$. Each of the input modes is computationally forward propagated, while each output mode is backwards propagated to each of the phase layers comprising of phase planes $\{P_1,P_2\}$ and \{$P_3,P_4\}$. The phase of each of these planes is then updated to minimize the phase difference between the wavefront of input and output modes. This process is repeated a few times until a convergence is observed. We then display the optimised phase-plane solutions $P_1$-$P_4$ in the physical setup for it to work as an optical circuit. 

\section{Programmable routing of entanglement over a single channel}
\label{sup_sec:routing}

To route $d$-dimensional entanglement (with $d=2,3$) using a single channel between pairs of nodes, we start by considering the truncated input entangled states of the form $|\Phi\rangle = \frac{1}{\sqrt{d}}\sum_{i=0}^{d-1}\ket{ii}$. The set $\{\ket{d}\}_d$ forms a $d$-dimensional macro-pixel basis. On the input side of the circuit, we select the macro-pixel modes $\{\ket{0}_{1},...,\ket{d}_{1},\ket{0}_{2},...,\ket{d}_{2}\}$, where the subscripts 1 and 2 indicate which input signal photon the modes correspond to. For the output modes, we select $d$ foci on each of the output ports of the circuit, which are directed to B and G. On A and H, single-outcome projective measurements on the $d$-dimensional macro-pixel basis over the idler photons herald the corresponding input modes.

The simultaneous distribution of qubit entanglement between AB and GH (AG and BH) is achieved with the 4x4 dimensional $\mathbb{I}$ ($\mathbb{X}$) gate.  We can also program 6x6 dimensional $\mathbb{I}$ ($\mathbb{X}$) operations to share three-dimensional entanglement between AB and GH, or AG and BH. 

\begin{figure*}[ht!]
    \centering
    \includegraphics[width=\textwidth]{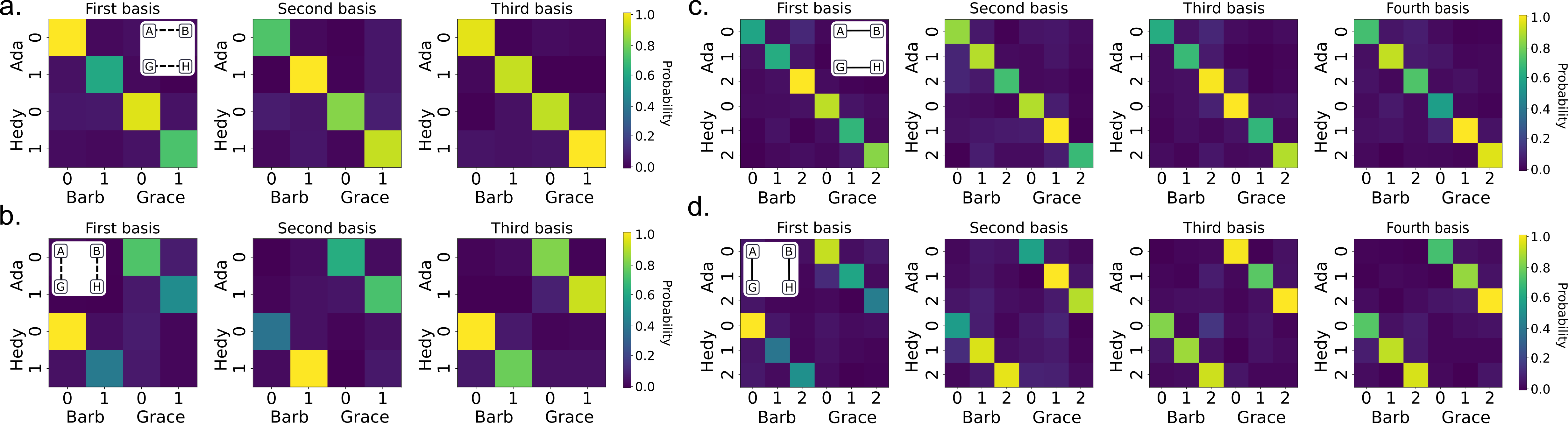}
    \caption{\textbf{Two-dimensional and three-dimensional entanglement routing}. We distribute entanglement between two users via a single channel either with a two-dimensional entangled state (a,b) or a three-dimensional state (c,d). We perform measurements in all mutually unbiased bases to characterise the operation of each configuration. The correlation matrices correspond to two-photon coincidence measurements, showcasing the circuit's operation.}
    \label{fig:3dimEntDist}
\end{figure*}

As described in~\ref{sec:ent_distribution}, we use two-fold coincidence measurements in all mutually unbiased bases (Fig.~\ref{fig:3dimEntDist}) to estimate fidelities to the maximally entangled state in dimensions $d=2,3$~\cite{Bavaresco:2018gw}. As shown in Table~\ref{table:Fid_singlechannel}, the obtained fidelities violate the dimensionality bounds, allowing us to certify the routing of two-dimensional and three-dimensional entanglement over the network. 

\begin{smallboxtable}{Fidelities to the maximally entangled state of the states shared in the programmable network using a single channel}{Fid_singlechannel}
\begin{center}
\begin{tabular}{|c|c|c|c|c|}
%\begin{tabular}{|m{8em}|m{8em}|m{8em}|m{8em}|m{8em}|}
    \hline
  \centering Gate & \multicolumn{2}{|c|}{2-dim states} & \multicolumn{2}{|c|}{3-dim states} \\
   \hline
  \multirow{2}{*}{$\mathbb{I}$} & \centering AB & \centering GH &\centering AB &GH\\
\cline{2-5}
                              & $86.2 \pm 1.1 \%$ & $83.1 \pm 1.1 \%$ & $73.1 \pm  1.8\%$ & $77.0 \pm 1.5 \%$\\
\hline
\centering\multirow{2}{*}{$\mathbb{X}$} & AG & BH & AG & BH\\
\cline{2-5}
                              & $81.2 \pm 1.2 \%$ & $83.6 \pm 1.2 \%$ & $77.4 \pm 1.8\%$ & $74.7 \pm 1.8\%$ \\
\hline
\centering\multirow{2}{*}{$\mathbb{SWAP}$} & \multicolumn{2}{c|}{AH} & \multicolumn{2}{c|}{-} \\
\cline{2-5} 
                                &  \multicolumn{2}{c|}{88.1 $\pm$ 2.0} & \multicolumn{2}{c|}{-}\\
\hline
\end{tabular}
\end{center}

\vspace*{10pt}
\footnotesize{*Errors are reported to three standard deviations}
\color{black}
\end{smallboxtable}

\section{Circuit stability}
\label{sup_sec:stability}

\begin{figure*}[ht!]
    \centering
    \includegraphics[width=0.7\textwidth]{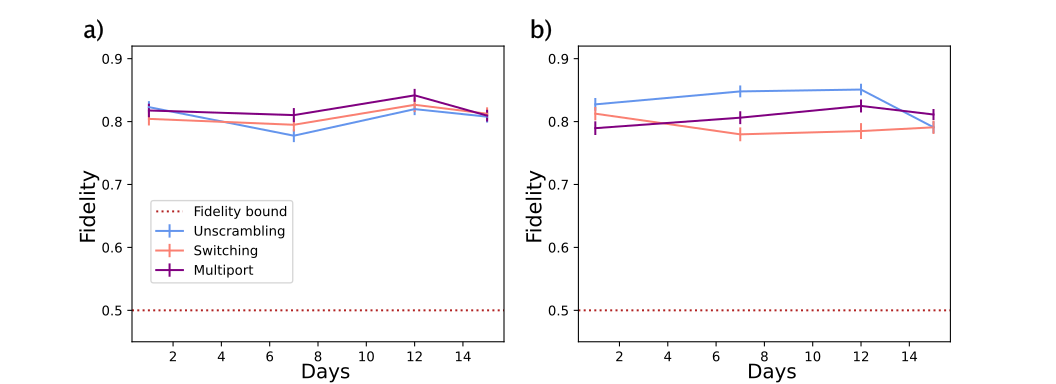}
    \caption{\textbf{Entanglement routing stability}. We measure the fidelities to the two-dimensional maximally entangled state of the states shared through the three different 8x8 unitary gates, implemented repeatedly over two weeks. Average fidelities over the two channels are shown for a) states shared with node Ada, and b) states shared with node Hedy. With $\mathbb{T}_I$, entanglement is either routed between AB and GH (Unscrambling), between AG and BH (Switching) with $\mathbb{T}_S$, or distributed from one party to the other two separate parties with $\mathbb{T}_M$ (Multiport). The red dotted line indicates the fidelity bound for a separable state.}
    \label{fig:stability}
\end{figure*}

We study the stability of the complex media-based circuit by characterising the performance of the programmable circuit over time. With a single initial characterisation of the mode-mixing in the MMF, and without performing any experimental realignment, we regularly test our highest-dimensional gates (8 $\times$ 8) over 14 days. Fidelity estimates of the states shared over the two channels are obtained through two-fold coincidence measurements on all mutually unbiased basis. As shown in Fig.~\ref{fig:stability}, the resulting fidelities, calculated with one standard deviation, consistently exceed the bound for a two-dimensional separable state. The consistent certification of the multiplexed and adaptable entanglement routing over time demonstrates the robustness of our platform.

\section{Entanglement swapping}
\label{sup_sec:swap}

An initial basic operation is to swap entanglement over one single channel per node. We consider two-dimensional input states of the form: $|\phi\rangle = \frac{1}{\sqrt{2}}\left( |00\rangle + |11\rangle \right)$, each produced by a different source. The complete initial four-photon state is then described as 

\begin{equation}
    \ket{\Phi} = \frac{1}{2}\left( |0000\rangle + |0101\rangle + |1010\rangle +|1111\rangle \right),
\end{equation}

where each element of the ket corresponds to the state of the photons $s_1,s_2,i_1,i_2$.

The entanglement swapping operation acting on signal photons $s_1$ and $s_2$ of $\ket{\Phi}$ is defined by the operation: 

\begin{equation}
\label{eq:swap_circuit_defn_4modes}
\mathbb{T}_S = \; 
\frac{1}{\sqrt{2}}
    \begin{bmatrix}
        1 & 0 & 1 & 0 \\
        0 & 1 & 0 & 1 \\
        1 & 0 & -1 & 0 \\
        0 & 1 & 0 & -1 \\
    \end{bmatrix}
\end{equation}

Following the transformation, we perform four-fold coincidence measurements between the output ports and the heralding photons, projecting the state shared between nodes A and H onto the maximally entangled target state $\ket{\Psi} = \left(\ket{01} + e^{i\varphi} \ket{10}\right)$, where $\varphi$ is a relative phase between the two sources introduced by the set-up.

This single-channel entanglement swapping protocol can be extended to operate across two simultaneous channels for connecting two pairs of remote users. In this scenario, the state of each source is expressed as $\ket{\phi} = \frac{1}{\sqrt{2}}\left(\ket{\phi}^{(\text{ch1})} + \ket{\phi}^{(\text{ch2})}\right)$, where  $\ket{\phi}^{(\text{ch1})} = \frac{1}{\sqrt{2}}\left(\ket{00} + \ket{11}\right)$ and $\ket{\phi}^{(\text{ch2})} = \frac{1}{\sqrt{2}}\left(\ket{22} + \ket{33}\right)$. Consequently, the initial four-photon state becomes a four-dimensional entangled state of the form:

\begin{equation}
    \ket{\Phi} = \frac{1}{2}\left(\ket{\phi}^{(\text{ch1})}_1+\ket{\phi}^{(\text{ch2})}_{1}\right)\otimes \left(\ket{\phi}^{(\text{ch1})}_2+\ket{\phi}^{(\text{ch2})}_{2}\right)
\end{equation}

where the subscript indicates the source. Then, the unitary operation that allows us to swap entanglement is given by Eq.~\ref{eq:swap_circuit_defn} in the main text. 

After the transformation of the signal photon states, we perform two sets of four-fold coincidence measurements across coincidences the two network channels, projecting the states shared between distant nodes A and H onto two different maximally entangled target states: $\ket{\Psi}^{\text{(ch1)}} = \frac{1}{\sqrt{2}}\left(\ket{01} - e^{i\varphi} \ket{10}\right)$ for the state shared between $A_1H_1$, and $\ket{\Psi}^{\text{(ch2)}} = \frac{1}{\sqrt{2}}\left(\ket{23} - e^{i\varphi} \ket{32}\right)$ for the state between $A_2H_2$. As before, $\varphi$ represents the relative phase introduced by the setup. This process allows us to simultaneously obtain two distinct swapped states between $A_1H_1$ and $A_2H_2$.

In both cases, the set of four-photon correlation measurements allows us to characterise swapped entangled states through quantum state tomography as detailed in ~\ref{sup_sec:QST}. We determine the relative phase and reconstruct the estimated states for each scenario and channel, achieving fidelities to the target state $\ket{\Psi}$ of more than $77\%$, thus successfully validating our protocol.

\section{Quantum State Tomography}
\label{sup_sec:QST}

Each realisation of the SWAP gate in the programmable quantum network generates a bipartite quantum state between A and H upon which we perform full quantum state tomography by measuring a tomographically complete set of measurements consisting of all local Pauli eigenvectors and performing Maximum-likelihood state estimation~\cite{Hradil2004}.
The complete set of measurements consists $\{\Pi^{AH}_{ambn} = \Pi^A_{a|m}\otimes \Pi^H_{b|n} \}_{abmn}$ where $\Pi^A_{a|m}=\ketbra{v_{a|m}}{v_{a|m}}$ and $a\in\{1,-1\}$ indexes the eigenvector of the Pauli matrix, $\sigma_m$ for $m\in\{X,Y,Z\}$. 
For each measurement, $n_{ambn}$ 4-photon coincidences are recorded and the maximum-likelihood estimator is obtained by iterating,
\be
\rho_{k+1}=\mathcal{N}\bigl[ \mathcal{R}(\rho_k)\rho_k \mathcal{R}(\rho_k)\bigr],
\ee
where
\be
\mathcal{R}(\rho):=\sum_{i\sim ambn}\tfrac{1}{M_{mn}}\frac{n_{i}}{\tr[\Pi^{AH}_{i}\rho]} \Pi^{AH}_{i},
\ee
with $\rho_0=\mathbb{I}/d$ taken as the maximally mixed state, $M_{mn}=\sum_{ab} n_{ambn}$ is the 4-photon events in each basis, and $\mathcal{N}[\rho]:=\rho\tr[\rho]$ imposes unit trace. To estimate the statistical noise present in our state estimates we perform Monte-Carlo bootstrapping by taking Poissonian samples from our data and repeating the estimation procedure. We obtain 2000 state estimates and their corresponding state fidelities to arrive at the 1STD standard error stated on our presented state fidelities.    

We perform quantum state tomography to recover the density matrix of the states shared users in nodes A and H to demonstrate the success of our entanglement swapping protocol. We estimate a fidelity to the two-dimensional maximally entangled target state of $88.1 \pm 2.0\%$ for single channel swapping. In the multiplexed case, we estimate fidelities of 77.1$\pm$ 3.3\% for the first channel ($A_1H_1$) and 83.2$\pm$ 2.7\% for the second channel ($A_2H_2$).